\documentclass[aps,pra,twocolumn,showpacs,superscriptaddress]{revtex4-1}%,showpacs,,preprint
\usepackage{amssymb}
\usepackage{txfonts}%and
\usepackage{graphicx}

\begin{document}

\title{Quantifying entanglement of arbitrary-dimensional multipartite pure states in terms of the singular values of coefficient matrices}

\emph{}
\author{Hui Li}
\thanks{These authors contributed equally to this work.}
\affiliation{State Key Laboratory of Low-Dimensional Quantum Physics and Department of Physics, Tsinghua University, Beijing 100084, China}
\author{Shuhao Wang}
\thanks{These authors contributed equally to this work.}
\affiliation{State Key Laboratory of Low-Dimensional Quantum Physics and Department of Physics, Tsinghua University, Beijing 100084, China}
\author{Jianlian Cui}
\affiliation{Department of Mathematical Sciences, Tsinghua University, Beijing 100084, China}
\author{Guilu Long}
\email{gllong@mail.tsinghua.edu.cn}
\affiliation{State Key Laboratory of Low-Dimensional Quantum Physics and Department of Physics, Tsinghua University, Beijing 100084, China}
\affiliation{Tsinghua National Laboratory for Information Science
and Technology, Beijing 100084, China}
\date{\today }

\begin{abstract}
The entanglement quantification and classification of multipartite quantum states are two important research fields in quantum information.
In this work, we study the entanglement of arbitrary-dimensional multipartite pure states by looking at the averaged partial entropies of various bipartite partitions of the system, namely, the so-called Manhattan distance ($l_1$ norm) of averaged partial entropies (MAPE), and it is proved to be an entanglement measure for pure states.
We connected the MAPE with the coefficient matrices, which are important tools in entanglement classification and reexpressed the MAPE for arbitrary-dimensional multipartite pure states by the nonzero singular values of the coefficient matrices.
The entanglement properties of the $n$-qubit Dicke states, arbitrary-dimensional Greenberger-Horne-Zeilinger states, and $D_3^n$ states are investigated in terms of the MAPE, and the relation between the rank of the coefficient matrix and the degree of entanglement is demonstrated for symmetric states by two examples.
\end{abstract}
\pacs{03.67.Mn, 03.65.Ud}

 \maketitle

\section{Introduction}
The entanglement of quantum systems was pointed out by Einstein, Podolsky, and Rosen (EPR) \cite{epr} and Schr{\" o}dinger \cite{schrodinger}.
Then the concept of entanglement was brought into the physical world with extraordinary properties and applications \cite{horodecki2009},
and it plays vital roles in quantum information theoretically and experimentally. Currently, entanglement is an essential resource for quantum information, which includes quantum teleportation, quantum cryptography, quantum computation, etc \cite{bennett1993,bennett2000,book}.
The study of quantum entanglement has become more and more popular with the explosive development of quantum information, two of the most important studies are the classification and the quantification of entanglement.

The main approach of entanglement classification under stochastic local operations and classical communication (SLOCC) is to find an invariant which is preserved under SLOCC,
and considerable research has been conducted since the beginning of this century \cite{vidal2000,verstraete2002,lamata2006,borsten2010,viehmann2011,chen2006,bastin2009,miyake2003,cornelio2006,linchen2006,eric2010}.
Recently, Li and Li have proposed the coefficient matrices as important tools in entanglement classification
under SLOCC \cite{dafali2012,dafali2012b}.
If we have an \emph{n}-qubit pure state $\left| \psi \right\rangle$, we can always expand $\left| \psi \right\rangle$ as $\left| \psi \right\rangle   = \sum\nolimits_{i = 0}^{{2^n} - 1} {{a_i}\left| i \right\rangle }$, where ${a_i}$ are the coefficients and $\left| i \right\rangle $ are the binary basis states. The coefficient matrices corresponding to $\left| \psi \right\rangle$ can be constructed as
\begin{widetext}
\begin{eqnarray}
C_{1\cdots l,l+1 \cdots n}(|\psi \rangle_{1\cdots n})=
\left( {\begin{array}{*{20}{c}}
{{a_{\underbrace {0 \cdots 0}_{l}\underbrace {0 \cdots 0}_{n-l}}}}& \cdots &{{a_{\underbrace {0 \cdots 0}_{l}\underbrace {1 \cdots 1}_{n-l}}}}\\
{{a_{\underbrace {0 \cdots 1}_{l}\underbrace {0 \cdots 0}_{n-l}}}}& \cdots &{{a_{\underbrace {0 \cdots 1}_{l}\underbrace {1 \cdots 1}_{n-l}}}}\\
 \vdots & \vdots & \vdots \\
{{a_{\underbrace {1 \cdots 1}_{l}\underbrace {0 \cdots 0}_{n-l}}}}& \cdots &{{a_{\underbrace {1 \cdots 1}_{l}\underbrace {1 \cdots 1}_{n-l}}}}
\end{array}} \right),
\end{eqnarray}
\end{widetext}
where $1 \le l \le n-1$.
As a natural generalization, an arbitrary-dimensional multipartite pure state $\left|\psi\right\rangle$ in the \emph{n}-partite
Hilbert space ${\cal H}^{n}={\cal H}_{1}\otimes{\cal H}_{2}\otimes\cdots\otimes{\cal H}_{n}$,
where ${\cal H}_{1},{\cal H}_{2},\cdots,{\cal H}_{n}$ have the dimensions
$d_{1},d_{2},. . .,d_{n}$, respectively, can be expanded as
\begin{eqnarray}
\left|\psi\right\rangle =\sum _{i=0}^{\prod\nolimits _{k=1}^{n}{d_{k}}-1}{{a_{i}}\left|{s_{1}}{s_{2}}\cdots{s_{n}}\right\rangle }, \end{eqnarray}
where ${a_{i}}$ are the coefficients and $\left|{s_{1}}{s_{2}}\cdots{s_{n}}\right\rangle $
are the basis states,
\begin{eqnarray}
\left|{{s_{1}}{s_{2}}\cdots{s_{n}}}\right\rangle =\left|{s_{1}}\right\rangle \otimes\left|{s_{2}}\right\rangle \otimes\cdots\otimes\left|{s_{n}}\right\rangle, 
\end{eqnarray}
with ${s_{k}}\in\{0,1,\cdots,{d_{k}-1}\},k=1,\cdots,n$.
We can construct the coefficient matrices by arranging the coefficients in a lexicographic ascending order \cite{wang2012}:
\begin{widetext}
\begin{eqnarray}
C_{1\cdots l,l+1 \cdots n}(|\psi \rangle_{1\cdots n})=
{\left({\begin{array}{ccc}
{a_{\underbrace{0\cdots0}_{l}\underbrace{0\cdots0}_{n-l}}} & \cdots & {a_{\underbrace{0\cdots0}_{l}\underbrace{{d_{n-l}}-1\cdots{d_{n}}-1}_{n-l}}}\\
{a_{\underbrace{0\cdots1}_{l}\underbrace{0\cdots0}_{n-l}}} & \cdots & {a_{\underbrace{0\cdots1}_{l}\underbrace{{d_{n-l}}-1\cdots{d_{n}}-1}_{n-l}}}\\
\vdots & \vdots & \vdots\\
{a_{\underbrace{{d_{1}}-1\cdots{d_{l}}-1}_{l}\underbrace{0\cdots0}_{n-l}}} & \cdots & {a_{\underbrace{{d_{1}}-1\cdots{d_{l}}-1}_{l}\underbrace{{d_{n-l}}-1\cdots{d_{n}}-1}_{n-l}}}\end{array}}\right)}.
\end{eqnarray}
\end{widetext}
Each permutation of qubits (or qudits) gives a permutation $\{q_1,q_2, \cdots ,q_n\}$ of $\{1,2, \cdots ,n\}$. So in this case, the coefficient matrices $C_{q_1 \cdots q_l,q_{l+1} \cdots q_n}(\left| \psi  \right\rangle )$ [$C_{q_1 \cdots q_l}(\left| \psi  \right\rangle )$ for short, omitting the column qudits] can be constructed by taking the corresponding permutation.
The coefficient matrices $C_{q_1 \cdots q_l}(\left| \psi  \right\rangle )$ have been proved to be invariant under SLOCC \cite{dafali2012,wang2012}, which provides us with an approach of entanglement classification for arbitrary-dimensional multipartite pure states.

Despite the classification of entanglement, the quantification of entanglement is also an important research area in quantum information. Much effort has been put into it in recent years \cite{bennett1996a,bennett1996b,vedral1997,vedral1998,wooters1998,uhlmann2000,cerf1997,peres1996,plenio2005}. However, the situation becomes much more complicated when faced with many particles.  Actually, only the simplest case, where states have two particles, can be completely described by current theories.
There are a variety of methods of entanglement quantification of multipartite states \cite{thapliyal1999,bennett2000p,schmid2008,gour2010,csyu2006,carvalho2004}.
In a quantum system with many particles, no particle is superior to the others; thus the calculation treat all the particles equally. Our entanglement measure, defined in this context, accounts for all the particles.

In this paper, we propose an entanglement measure named the Manhattan distance of averaged partial entropies (MAPE). The connection between the MAPE and the coefficient matrices is established. By means of the MAPE, we discover many noble entanglement properties of several arbitrary-dimensional multipartite pure states. With two examples, we show that the rank of $C_{q_1 \cdots q_{[n/2]}}$ and the degree of entanglement are  closely linked.

This paper is organized as follows: in Sec. \ref{MAP} we introduce an entanglement measure named the MAPE. The mathematical connection between the MAPE and the coefficient matrices is established. We prove that the MAPE is an entanglement measure for pure states. In Sec. \ref{APP} we investigate entanglement properties of the $n$-qubit Dicke states, arbitrary-dimensional Greenberger-Horne-Zeilinger (GHZ) states, and $D_3^n$ states in terms of the MAPE. The relation between the rank of $C_{q_1 \cdots q_{[n/2]}}$ and the MAPE, \emph{i.e.}, the degree of entanglement, is investigated for symmetric states using examples. In Sec. \ref{CON} we give a short summary and prospects.

\section{The MAPE and coefficient matrices}
\label{MAP}

Different from the bipartite partial entropy or its modified versions, the averaged partial entropies (APE) take into account all the partitions for a multipartite pure state.
The complete entanglement measure in terms of the APE was pointed out in Ref. \cite{dliu2010}, where they named the entanglement measure multiple entropy measures (MEMS).
Suppose $\{q_1,q_2, \cdots ,q_n\}$ is a permutation of $\{1,2, \cdots ,n\}$; the MEMS for multipartite pure
quantum states is defined as a vector,
\begin{equation}
\vec{S}=(S_1,S_2,\cdots,S_{[n/2]}),
\end{equation}
the elements of which are the APE,
\begin{equation}
S_l=\left[\prod_{q_1,\cdots ,q_l=1}^n E_{q_1,\cdots ,q_l}\right]^{1\over C_n^l},
\end{equation}
where $1 \le l \le [n/2]$,
$ E_{q_1,\cdots ,q_l}=-{\rm Tr}(\rho_{q_1\cdots
q_l}\log_2 \rho_{q_1\cdots q_l})
$
is the reduced von Neumann entropy with the
other $n-l$ particles being traced out, and
\begin{equation}
C_n^l={n!\over (n-l)! l!}.
\end{equation}

Our entanglement measure is defined as the Manhattan distance ($l_1$ norm) of APE (MAPE),
namely,
\begin{eqnarray}
\mathcal{M}&=&|S_1|+|S_2|+\cdots +|S_{[n/2]}|\nonumber\\
&=&S_1+S_2+\cdots +S_{[n/2]},
\end{eqnarray}
where we have considered that $S_l \ge 0$.
It needs to be noted that the $l_2$ norm of the APE cannot be used to define a measure since it is not an entanglement monotone, the proof of which is given in the Appendix.
We show that the MAPE is closely connected to the coefficient matrices; the relationship directly links entanglement quantification with entanglement classification.

\emph{Theorem 1}.
The MAPE of an arbitrary-dimensional multipartite pure state can be reexpressed by the nonzero singular values of the coefficient matrices, namely,
\begin{eqnarray}
&&\mathcal{M}=\nonumber\\
&&{\sum\nolimits_{l = 1}^{[n/2]} {{{\left[ {\prod\limits_{{q_1}, \cdots ,{q_l} = 1}^n  -  \sum\limits_i {\lambda _{{q_1} \cdots {q_l},i}^2{{\log }_2}\lambda _{{q_1} \cdots {q_l},i}^2} } \right]}^{\frac{1}{{C_n^l}}}}} },
\label{MAPE}
\end{eqnarray}
where $\lambda_{q_1 \cdots q_{l} ,i}$ are the nonzero singular values of $C_{q_{1}\cdots q_{l}}(|\psi \rangle_{1\cdots n}) $.

\emph{Proof.}
The relation between all the reduced density matrices and the coefficient matrices is given by \cite{dafali2012b}
\begin{eqnarray}
\rho _{q_{1}\cdots q_{l}}(|\psi \rangle_{1\cdots n})=
C_{q_{1}\cdots q_{l}}(|\psi \rangle_{1\cdots n}) C_{q_{1}\cdots
q_{l}}^{\dagger }(|\psi \rangle_{1\cdots n}), \label{relation}
\end{eqnarray}
where $C_{q_{1}\cdots q_{l}}^{\dagger }(|\psi \rangle_{1\cdots n})$ is the conjugate transpose of $C_{q_{1}\cdots q_{l}}(|\psi \rangle_{1\cdots n})$.

In the following proof, under the premise of no confusion, we use $\rho$ and $C$ to represent the reduced density matrix $\rho_{q_{1}\cdots q_{l}}(|\psi \rangle_{1\cdots n})$ and the coefficient matrix $C_{q_{1}\cdots q_{l}}(|\psi \rangle_{1\cdots n})$, respectively, for convenience.

The singular value decomposition of $C^\dagger$ is given by
\begin{equation}
C^\dagger=V \Sigma^\dagger U^\dagger.
\end{equation}
According to Eq. (\ref{relation}), the reduced density matrix can be expressed as
\begin{equation}
\rho  = C{C^\dag } = U\Sigma {V^\dag }V{\Sigma ^\dag }{U^\dag }.
\end{equation}
Since $V$ is unitary, namely, $V^\dag V=I$, we have
\begin{equation}
\rho  =  U\Sigma {\Sigma ^\dag }{U^\dag },
\end{equation}
which represents the diagonalization of $\rho$. The columns of $U$ are the eigenvectors of $\rho$.
Suppose the nonzero eigenvalues of $\rho$ are $\Lambda_i$; then
\begin{equation}
\Lambda_i=\lambda_i {\lambda_i}^*= {{\lambda _i}} ^2,
\label{lambdarelation}
\end{equation}
where $\lambda_i$ are the corresponding nonzero singular values of $C$.
The von Neumann entropy of $\rho$ is defined by
\begin{equation}
S(\rho)=-{\rm Tr}(\rho \rm{log_2} \rho).
\label{neumann}
\end{equation}
Equation (\ref{neumann}) can be reexpressed by the nonzero eigenvalues of $\rho$:
\begin{equation}
S(\rho)= - \sum\limits_i {{\Lambda _i}\log_2 {\Lambda _i}}.
\label{reexpress}
\end{equation}
Thus the von Neumann entropy of $\rho_{q_{1},\cdots, q_{l}}(|\psi \rangle_{1\cdots n})$ can be expressed as
\begin{eqnarray}
E(\rho_{q_{1},\cdots, q_{l}}(|\psi \rangle_{1\cdots n}) )=
 - \sum\limits_i {{{\lambda^2_{q_1 \cdots q_l ,i} }}\log_2 {{\lambda^2_{q_1 \cdots q_l ,i}}}},
\label{vonneumann}
\end{eqnarray}
where $\lambda_{q_1 \cdots q_l ,i}$ are the nonzero singular values of $C_{q_{1}\cdots q_{l}}(|\psi \rangle_{1\cdots n}) $.
Then we have
\begin{eqnarray}
S_l=
\left[\prod_{q_1,\cdots ,q_l=1}^n  - \sum\limits_i {{{\lambda^2_{q_1 \cdots q_l ,i} }}\log_2 {{\lambda^2_{q_1 \cdots q_l ,i}}}}\right]^{1\over C_n^l}.
\label{sl}
\end{eqnarray}

Therefore we get Eq. (\ref{MAPE}).

\emph{Theorem 2}. The MAPE is an entanglement measure for pure states.

\emph{Proof.}
We first prove that the MAPE is an entanglement monotone; namely, it does not increase, on average, under local operations and classical communication (LOCC). By using LOCC, a pure state $|\psi \rangle$ can be transformed into the state
\begin{equation}
|\phi_k \rangle=\frac{L_k |\psi \rangle}{\sqrt{{\rm Tr}{(L_k^\dagger L_k |\psi \rangle \langle \psi |)}}},
\end{equation}
with a probability $p_k$. Here $L_k = A_1^k \otimes \cdots \otimes A_n^k$ satisfies
\begin{equation}
\sum_k L_k^\dagger L_k = I,
\label{nom}
\end{equation}
Where $A_1^k,\cdots,A_n^k$ are local operators on each particle, with ${A_i^k}^\dagger A_i^k \le I$ $(i=1,\cdots,n)$, and $p_k={\rm Tr}(L_k^\dagger L_k |\psi \rangle \langle \psi |)$ represents the probability of obtaining $|\phi_k \rangle$ after LOCC \cite{book,vidal2000}. It can be easily verified that $\sum\limits_k p_k = 1$ according to Eq. (\ref{nom}). Noting that the von Neumann entropy does not increase on average under LOCC; thus the APE do not, on average, increase under LOCC, namely,
\begin{equation}
\sum\limits_k {{p_k}{S_l}(\left| {{\phi _k}} \right\rangle )} \le {S_l}(\left| \psi  \right\rangle ),
\end{equation}
it can be verified that
\begin{eqnarray}
\sum\limits_k {{p_k}\mathcal{M}(\left| {{\phi _k}} \right\rangle )} 
&=& \sum\limits_k {{p_k}[{S_1}(\left| {{\phi _k}} \right\rangle ) + {S_2}(\left| {{\phi _k}} \right\rangle ) +  \cdots  }\nonumber\\
&&+ {S_{[n/2]}}(\left| {{\phi _k}} \right\rangle )]
\le \mathcal{M}(\left| \psi  \right\rangle ).
\end{eqnarray}
Therefore the MAPE does not increase, on average, under LOCC.

It is easy to see that the MAPE is non-negative. Next, we prove the MAPE is zero for fully separable pure states.
For fully separable pure states, the ranks of all the coefficient matrices are 1 \cite{dafali2012b,wang2012q}.
Since
\begin{equation}
{\rm Tr}[\rho_{q_{1}\cdots q_{l}}(|\psi \rangle_{1\cdots n})]=1,
\end{equation}
we have
\begin{equation}
\sum\limits_i {{\lambda _{{q_1} \cdots {q_l},i}}^2}  = 1.
\end{equation}
In the case where $r(C_{q_{1}\cdots q_{l}}(|\psi \rangle_{1\cdots n}))=1$, there exists only one nonzero singular value 1.
According to Eq. (\ref{MAPE}), $\mathcal{M} = 0$.

Therefore the MAPE satisfies the requirements of an entanglement measure for pure states.

\emph{Theorem 3}.
For a genuinely entangled multipartite pure state, the MAPE is not zero.

\emph{Proof.}
It has been proved that a multipartite pure state is genuinely entangled if and
only if the ranks of all the coefficient matrices are greater than 1 \cite{dafali2012b,wang2012q},
which indicates that all $\lambda_{q_1 \cdots q_l ,i}$'s are not 1.
Thus $\mathcal{M} \neq 0$ for genuinely entangled pure states.

\section{Applications}
\label{APP}

In terms of the MAPE, we discuss the entanglement properties of the $n$-qubit Dicke states, arbitrary-dimensional GHZ states, and $D_3^n$ states.

\subsection{The $n$-qubit Dicke states}

The $n$-qubit Dicke states are defined as
\begin{eqnarray}
\left| {l_1,n} \right\rangle  =
{\left( {\frac{{n!}}{{{l_1}!{l_0}!}}} \right)^{ - 1/2}}\sum\limits_k {{P_k}\left| {\underbrace {1, \cdots ,1}_{{l_1}},\underbrace {0, \cdots ,0}_{{l_0}}} \right\rangle },
\end{eqnarray}
where $\left|1\right\rangle $ represents the excitation with respect to
the ground state $\left|0\right\rangle $,  $l_1$
is the number of excitations $\left|1\right\rangle $, which satisfy $0 \le l_1 \le n$ and $l_0=n-l_1$.
$\{{P_{k}}\}$ is the set of all permutations.

The MAPE of three-, six-, and nine-qubit Dicke states is shown in Fig. 1, which
indicates that the states are maximumly entangled when the energy levels $\left|0\right\rangle $ and $\left|1\right\rangle $ are equally occupied.

\begin{figure}[!h]
%[tpb]

\begin{centering}
\includegraphics[width=7.2cm]{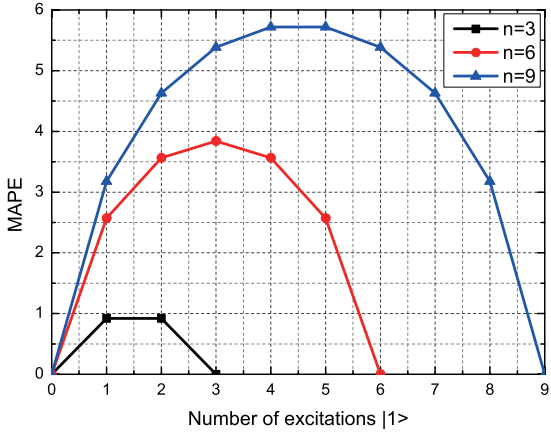}
\caption{(Color online) The MAPE of three, six and nine-qubit Dicke states.}
\par\end{centering}

\centering{}
\end{figure}

\subsection{The arbitrary-dimensional GHZ states}

The $n$-partite and $d$-dimensional GHZ state has a simple expression;
\begin{equation}
\left| {\rm GHZ} \right\rangle  = \frac{1}{{\sqrt d }}\sum\limits_{i = 0}^{d - 1} {\left| {\underbrace {ii \ldots i}_n} \right\rangle }.
\end{equation}
It can be calculated that all the coefficient matrices have the form
\begin{equation}
C=\left( {\begin{array}{*{20}{c}}
{\frac{1}{{\sqrt d }}}&0& \cdots &0&0\\
0& \ddots & \cdots &0&0\\
 \vdots & \vdots &{\frac{1}{{\sqrt d }}}& \vdots & \vdots \\
0&0& \cdots & \ddots &0\\
0&0& \cdots &0&{\frac{1}{{\sqrt d }}}
\end{array}} \right),
\end{equation}
where the coefficient matrices are usually not square matrices. They have $d$ diagonal element that are nonzero, and the nondiagonal elements are all zero.
Thus, for an $n$-partite and $d$-dimensional GHZ state, the coefficient matrices have $d$ nonzero singular values which equal to $\frac{1}{{\sqrt d }}$.
Therefore
\begin{equation}
{S_l} =  - {\log _2}\frac{1}{d}=\log_2 d,
\end{equation}
which obviously leads to
\begin{equation}
\mathcal{M}=[n/2]{\log _2}d.
\end{equation}

\subsection{The $D_3^n$ states}

The $D_3^n$ states are defined as
\begin{eqnarray}
\left|{{l_{1}},{l_{2}},n}\right\rangle =
{\left({\frac{{n!}}{{{l_{1}}!{l_{2}}!{l_{0}}!}}}\right)^{-\frac{1}{2}}}\sum\limits _{k}{{P_{k}}\left|{\underbrace{1,\cdots,1}_{{l_{1}}},\underbrace{2,\cdots,2}_{{l_{2}}},\underbrace{0,\cdots,0}_{{l_{0}}}}\right\rangle },\nonumber\\
\label{d3}
\end{eqnarray}
where $\left|1\right\rangle$ and $\left|2\right\rangle $ are the excitations,
$\left|0\right\rangle $ represents the ground state, and ${l_{0}},{l_{1}},{l_{2}}$
are the numbers of states $\left|0\right\rangle ,\left|1\right\rangle ,\left|2\right\rangle $,
respectively, which satisfy $0 \le {l_{1}}+{l_{2}} \le n$ and $l_0=n-l_1-l_2$. $\{{P_{k}}\}$
is the set that contains all permutations. The MAPE for $D_3^9$ states are shown in Fig. 2.
The result shows that $D_3^9$ states are maximumly entangled when $l_0=l_1=l_2=3$, namely, when the energy levels are equally occupied.

\begin{figure}[!h]
%[tpb]

\begin{centering}
\includegraphics[width=7.2cm]{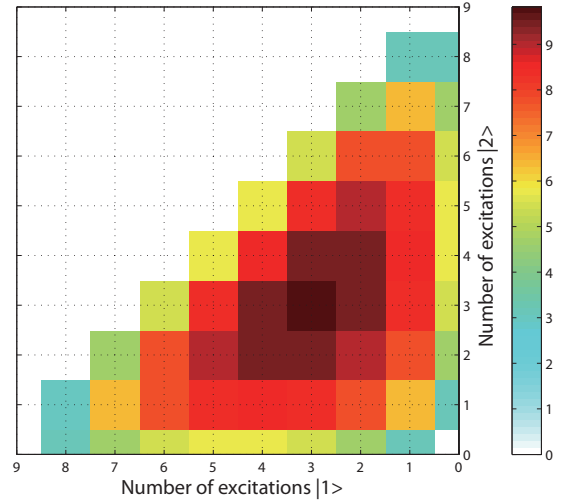}
\caption{(Color online) The MAPE of $D_3^9$ states.}

\par\end{centering}

\centering{}
\end{figure}

\subsection{Relation between the rank of $C_{q_{1},\cdots, q_{[n/2]}}$ and degree of entanglement}

The relation between the ranks of the coefficient matrices and degree of entanglement is of great interest \cite{wang2012}. The question is demonstrated for symmetric states by two examples, namely, the eight-qubit Dicke state and $D_3^9$ states.

It has been shown that the rank of $C_{q_{1},\cdots, q_{[n/2]}}(|\psi \rangle_{1\cdots n})$ corresponding to $n$-qubit Dicke states is $k+1$ (when $0 \le k \le [n/2]$) \cite{dafali2012}.
The rank of $C_{q_{1},\cdots, q_{[n/2]}}(|\psi \rangle_{1\cdots n})$ and $\mathcal{M}$ of eight-qubit Dicke states are shown in Fig. 3.
It can be seen that the rank of $C_{q_{1},\cdots, q_{[n/2]}}(|\psi \rangle_{1\cdots n})$ and $\mathcal{M}$ of eight-qubit Dicke states have the same trend, and the case where the rank of $C_{q_{1},\cdots, q_{[n/2]}}(|\psi \rangle_{1\cdots n})$ of eight-qubit Dicke states is maximized corresponds to the maximum degree of entanglement.

\begin{figure}[!h]
%[tpb]

\begin{centering}
\includegraphics[width=7.2cm]{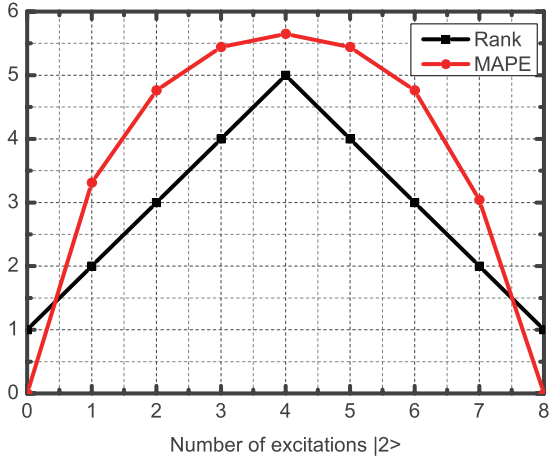}
\caption{(Color online) Rank of the coefficient matrix $C_{q_{1},\cdots, q_{[n/2]}}(|\psi \rangle_{1\cdots n})$ and the MAPE of eight-qubit Dicke states.}

\par\end{centering}

\centering{}
\end{figure}

Next, we study the $D_3^9$ states. Numerical results have shown that the rank of $C_{q_{1},\cdots, q_{[n/2]}}(|\psi \rangle_{1\cdots n})$ and $\mathcal{M}$ for $D_3^9$ are maximized simultaneously when $l_1,l_2,l_0$ are all 3.
The results for  $l_1$ fixed to 3 are shown in Fig. 4, which shows that the rank of $C_{q_{1},\cdots, q_{[n/2]}}(|\psi \rangle_{1\cdots n})$ is closely linked to the degree of entanglement.

\begin{figure}[!h]
%[tpb]

\begin{centering}
\includegraphics[width=7.2cm]{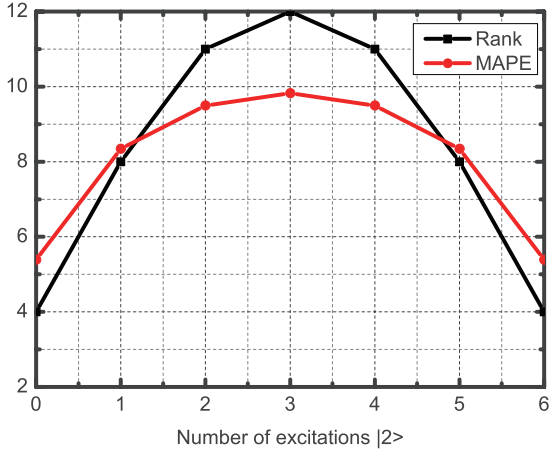}
\caption{(Color online) Rank of the coefficient matrix $C_{q_{1},\cdots, q_{[n/2]}}(|\psi \rangle_{1\cdots n})$ and the MAPE of $D_3^9$ states when $l_1=3$.}

\par\end{centering}

\centering{}
\end{figure}

\section{Conclusion}
\label{CON}

In summary, we have proposed an entanglement measure named the MAPE, and the mathematical connection between the MAPE and the coefficient matrices was established, which indicates that entanglement classification and quantification are closely linked to the number and value of the nonzero singular values of the coefficient matrices.
Examples were discussed to show that the MAPE is capable of dealing with quantum pure states with arbitrary dimensions. The rank of the coefficient matrix $C_{q_{1},\cdots, q_{[n/2]}}$ and the degree of entanglement for eight-qubit Dicke and $D_3^9$ states are proved to have positive correlations.

It needs to be noted that Eq. (\ref{vonneumann}) provides us with a way of calculating the von Neumann entropy in terms of the coefficient matrix, which is also a useful tool in analyzing other problems in a simpler manner. For instance, by means of the coefficient matrix and the criteria shown in Refs. \cite{bennett2000p,thapliyal1999}, it can be easily proved that for $n$-qubit Schmidt decomposable pure states, the ranks, \emph{i.e.},
the number of nonzero singular values, of the coefficient matrices $C_{q_1}(|\psi \rangle_{1\cdots n})$ are equal (being either 1 or 2), and in the case where the ranks $C_{q_1}(|\psi \rangle_{1\cdots n})$ are 2,
two nonzero singular values are one-to-one correspondent.
In the mean-time,  it can be proved that the $(n-1)$-partite reduced states $\rho _{{q_2} \cdots {q_n}}{(|\psi \rangle _{1 \cdots n}})$ of an $n$-qubit Schmidt decomposable state are all pure or mixed.

We expect that
our work could come up with further theoretical and
experimental results.

\section*{Acknowledgments}

This work was supported by the National Natural Science Foundation
of China (Grants No.11175094 and No.11271217) and the National Basic Research Program
of China (Grants No.2009CB929402 and No. 2011CB9216002).

\renewcommand{\theequation}{A.\arabic{equation}}
  % redefine the command that creates the equation no.
  \setcounter{equation}{0}  % reset counter 
\section*{Appendix}
We prove the monotonicity of $S_l$ $(1 \le l \le [n/2])$ cannot guarantee the monotonicity of the $l_2$ norm (module) of the APE, namely,
\begin{equation}
\label{mape2}
\mathcal{M}^{'}=|\vec{S}|=\sqrt{S_1^2+S_2^2+\cdots+S_{[n/2]}^2}.
\end{equation}

A mathematical counterexample can be given. Consider a pure state with $n=6$; without loss of generality, suppose the LOCC gives $|\psi \rangle \to {p_1}|{\phi _1}\rangle \left\langle {{\phi _1}} \right| + {p_2}|{\phi _2}\rangle \left\langle {{\phi _2}} \right|$. The monotonicity of the APE implies that
\begin{widetext}
\begin{eqnarray}
{p_1}{S_1}(|{\phi _1}\rangle ) + {p_2}{S_1}(|{\phi _2}\rangle ) \le {S_1}(\left| \psi  \right\rangle ),
{p_1}{S_2}(|{\phi _1}\rangle ) + {p_2}{S_2}(|{\phi _2}\rangle ) \le {S_2}(\left| \psi  \right\rangle ),
{p_1}{S_3}(|{\phi _1}\rangle ) + {p_2}{S_3}(|{\phi _2}\rangle ) \le {S_3}(\left| \psi  \right\rangle ).
\end{eqnarray}
Since both sides of the inequalities are non-negative, further calculation yields
\begin{eqnarray}
\label{mono}
{\left[ {{p_1}{S_1}(|{\phi _1}\rangle ) + {p_2}{S_1}(|{\phi _2}\rangle )} \right]^2} \le S_1^2(\left| \psi  \right\rangle ),
{\left[ {{p_1}{S_2}(|{\phi _1}\rangle ) + {p_2}{S_2}(|{\phi _2}\rangle )} \right]^2} \le S_2^2(\left| \psi  \right\rangle ),
{\left[ {{p_1}{S_3}(|{\phi _1}\rangle ) + {p_2}{S_3}(|{\phi _2}\rangle )} \right]^2} \le S_3^2(\left| \psi  \right\rangle ).
\end{eqnarray}
It can be calculated that
\begin{eqnarray}
{[{p_1}\mathcal{M}^{'}(\left| {{\phi _1}} \right\rangle ) + {p_2}\mathcal{M}^{'}(\left| {{\phi _2}} \right\rangle )]^2} 
&=& ({p_1}\sqrt {S_1^2(\left| {{\phi _1}} \right\rangle ) + S_2^2(\left| {{\phi _1}} \right\rangle ) + S_3^2(\left| {{\phi _1}} \right\rangle )}  + {p_2}\sqrt {S_1^2(\left| {{\phi _2}} \right\rangle ) + S_2^2(\left| {{\phi _2}} \right\rangle ) + S_3^2(\left| {{\phi _2}} \right\rangle )} )^2\nonumber\\
&=& p_1^2[S_1^2(\left| {{\phi _1}} \right\rangle ) + S_2^2(\left| {{\phi _1}} \right\rangle ) + S_3^2(\left| {{\phi _1}} \right\rangle )] + p_2^2[S_1^2(\left| {{\phi _2}} \right\rangle ) + S_2^2(\left| {{\phi _2}} \right\rangle ) + S_3^2(\left| {{\phi _2}} \right\rangle )]\nonumber\\
&& + 2{p_1}{p_2}\sqrt {S_1^2(\left| {{\phi _1}} \right\rangle ) + S_2^2(\left| {{\phi _1}} \right\rangle ) + S_3^2(\left| {{\phi _1}} \right\rangle )}   \sqrt {S_1^2(\left| {{\phi _2}} \right\rangle ) + S_2^2(\left| {{\phi _2}} \right\rangle ) + S_3^2(\left| {{\phi _2}} \right\rangle )}.
\end{eqnarray}
Note that
\begin{eqnarray}
\sqrt {S_1^2(\left| {{\phi _1}} \right\rangle ) + S_2^2(\left| {{\phi _1}} \right\rangle ) + S_3^2(\left| {{\phi _1}} \right\rangle )}   \sqrt {S_1^2(\left| {{\phi _2}} \right\rangle ) + S_2^2(\left| {{\phi _2}} \right\rangle ) + S_3^2(\left| {{\phi _2}} \right\rangle )}
\ge  {S_1}(\left| {{\phi _1}} \right\rangle ){S_1}(\left| {{\phi _2}} \right\rangle ) + {S_2}(\left| {{\phi _1}} \right\rangle ){S_2}(\left| {{\phi _2}} \right\rangle ) + {S_3}(\left| {{\phi _1}} \right\rangle ){S_3}(\left| {{\phi _2}} \right\rangle );\nonumber\\
\end{eqnarray}
we further get
\begin{eqnarray}
{[{p_1}\mathcal{M}^{'}(\left| {{\phi _1}} \right\rangle ) + {p_2}\mathcal{M}^{'}(\left| {{\phi _2}} \right\rangle )]^2} 
\ge  {[{p_1}{S_1}(\left| {{\phi _1}} \right\rangle ) + {p_2}{S_1}(\left| {{\phi _2}} \right\rangle )]^2}  + {[{p_1}{S_2}(\left| {{\phi _1}} \right\rangle ) + {p_2}{S_2}(\left| {{\phi _2}} \right\rangle )]^2}  + {[{p_1}{S_3}(\left| {{\phi _1}} \right\rangle ) + {p_2}{S_3}(\left| {{\phi _2}} \right\rangle )]^2}.
\end{eqnarray}
Recall that
\begin{eqnarray}
&& \mathcal{M}^{'2}(\left| {{\psi}} \right\rangle ) = S_1^2(\left| \psi  \right\rangle ) + S_2^2(\left| \psi  \right\rangle ) + S_3^2(\left| \psi  \right\rangle ).
\end{eqnarray}
Therefore, according to Eq. (\ref{mono}), the monotonicity of $\mathcal{M}^{'}$ given in Eq. (\ref{mape2}) is not guaranteed by the monotonicity of $S_1$, $S_2$ and $S_3$.
\end{widetext}

\end{document}